\documentclass[%
 reprint,
 amsmath,amssymb,
 aps,
]{revtex4-1}

\usepackage{graphicx}
\usepackage{dcolumn}
\usepackage{bm}
\usepackage{color}
\usepackage[dvipdfmx]{hyperref}

\begin{document}

\preprint{APS/123-QED}

\title{Tunable Majorana corner modes in noncentrosymmetric superconductors: \\ Tunneling spectroscopy and edge imperfections}

\author{S. Ikegaya$^{1}$, W. B. Rui$^{2}$, D. Manske$^{1}$ and Andreas P. Schnyder$^{1}$}
\affiliation{$^{1}$Max-Planck-Institut f\"ur Festk\"orperforschung, Heisenbergstrasse 1, D-70569 Stuttgart, Germany\\
$^{2}$ Department of Physics and HKU-UCAS Joint Institute for Theoretical and Computational Physics at Hong Kong,
The University of Hong Kong, Pokfulam Road, Hong Kong, China}

\date{\today}

\begin{abstract}
Majorana corner modes appearing in two-dimensional second-order topological superconductors
have great potential applications for fault-tolerant topological quantum computations.
We demonstrate that in the presence of an in-plane magentic field two-dimensional ($s+p$)-wave superconductors host Majorana corner modes,
whose location can be manipulated by the direction of the magnetic field.
In addition, we discuss the effects of edge imperfections on the Majorana corner modes.
We describe how different edge shapes and edge disorder affect the number and controllability of the Majorana corner modes,
which is of relevance for the implementation of topological quantum computations.
We also discuss tunneling spectroscopy in the presence of the Majorana corner modes, where a lead-wire is attached to the corner of the noncentrosymmetric superconductor.
The zero-bias differential conductance shows a distinct periodicity with respect to the direction of the magnetic field,
which demonstrates the excellent controllability of the Majorana corner modes in this setup. 
Our results lay down the theoretical groundwork for observing and tuning Majoran corner modes in experiments on ($s+p$)-wave superconductors.  
\end{abstract}

\maketitle

\section{Introduction}
A central subject in physics of topological superconductivity~\cite{kane_10,zhang_11,tanaka_12r,sato_17}
is the realization of fault-tolerant topological quantum computations~\cite{ivanov_01,kitaev_03,sarma_08,fisher_11,sarma_15}
using Majorana zero modes~\cite{green_00,kitaev_01,wilczek_09}.
So far, the existence of the Majorana zero modes has been experimentally demonstrated in various systems such as
semiconductor-superconductor hybrids
\cite{sarma_10,oreg_10,kouwenhoven_12,deng_12,kouwenhoven_18,halperin_17,haim_19,setiawan_19,ikegaya_20(2),nichele_19,yacoby_19,marcus_20},
magnetic atom chains on superconductors~\cite{beenakker_11,yazdani_13,yazdani_14,yazdani_17},
and superconducting topological insulators~\cite{kane_08,gao_18,ando_11,tanaka_12,zhang_10,wang_17}.
However, to achieve the topological quantum computations, we have to manipulate the position of the Majorana zero modes for performing braiding operations,
which is still a challenging task at the current stage.

Recently, a fresh route for realizing topological quantum computations has been discussed in the context of higher-order topological superconductivity
\cite{balents_15,hughes_17,fang_17,brouwer_17,khalaf_18,hughes_18,wang_18,zhang_18,nori_18,zhu_19,sarma_19,yan_19,zhang_20,volovik_10,zhu_18,klinovaja_19,
liu_19a,trauzettel_20a,soluyanov_20,trauzettel_20}:
a $n$th-order topological superconductor in $d$ dimensions can host Majorana zero modes in $(d-n)$ dimensions for $n \geq 2$,
whereas conventional topological superconductors correspond to the case with $n=1$.
Although the study of higher-order topological superconductors is still in its infancy,
there have already been several theoretical predictions for 
two-dimensional second-order topological superconductors in which the position of emergent Majorana corner modes (MCMs) can be controlled by
varying experimentally tunable parameters~\cite{volovik_10,zhu_18,klinovaja_19,liu_19a,trauzettel_20a,soluyanov_20,trauzettel_20}.
Moreover, braiding operations using the advantages of such tunable MCMs have been demonstrated theoretically~\cite{trauzettel_20a,soluyanov_20,trauzettel_20}.
Thus, this recent development in the physics of tunable MCMs shows great promise for the realization of fault-tolerant topological quantum computations.

A fundamental strategy  for obtaining tunable MCMs is the application of magnetic fields to time-reversal invariant topological superconductors,
which host helical Majorana edge states~\cite{brouwer_17,khalaf_18,volovik_10}.
On the basis of this strategy, the presence of tunable MCMs has been demonstrated in
Rashba bilayers coupled to superconductors~\cite{klinovaja_19},
semiconductor/iron-based superconductor hybrids~\cite{zhu_18},
as well as spin-triplet helical $p$-wave superconductors~\cite{zhu_18,soluyanov_20}.
Even so, since these proposals have not yet been implemented experimentally, it is important to continue to propose alternative models hosting tunable MCMs.
Thus, as the first objective of this paper, we demonstrate that a two-dimensional noncentrosymmetric ($s+p$)-wave superconductor
\cite{fujimoto_07,agterberg_07,sigrist_07,nagaosa_09,sato_09,sato_10,schnyder_15},
which is experimentally realized, e.g., in
CePt$_3$Si \cite{rogl_04,onuki_05,bauer_05,haller_09,sigrist_04,fujimoto_05,sigrist_06,sigrist_07(2)}
or in CeRhSi$_3$~\cite{terashima_07,onuki_08,fujimoto_08,thalmeier_09},
can harbor the tunable MCMs by applying in-plane magnetic fields.

Moreover, while most previous studies on tunable MCMs consider square superconducting islands with clean edges,
experimental implementations will most certainly deviate from such an ideal geometry.
Thus, as the second objective of this paper, we study the effects of edge imperfections on the MCMs.
For a square geometry with clean edges, we can find two stable Majorana zero modes.
However, for more complicated edge configurations, the number of MCMs becomes more than two and depends on the applied magnetic field direction. 
In this case, we can no longer find a stable pair of Majorana zero modes that can make a full circle around the system, which complicates the braiding process.
Thus, the edge configuration is an important factor for accomplishing the topological quantum computations in experiments.

At present, to the best of our knowledge, there are no specific theoretical proposals for detecting the tunable MCMs.
Thus, as the third objective of this paper, we study the tunneling spectroscopy in the presence of the MCMs,
where the normal lead wire is attached to the vicinity of the corner of the noncentrosymmetric superconductor.
We demonstrate that the zero-bias conductance becomes a periodic function with respect to the direction of the applied magnetic field.
This characteristic periodicity is understood by the fact that
the zero-bias conductance is enhanced only when the MCM exits at the corner connected with the lead wire.
As a result, we propose a smoking-gun experiment for the detection of controllable MCMs.

The organization of this paper is as follows.
In Sec.~\ref{sec:2}, we first derive an effective edge theory describing the emergence of tunable MCMs in a noncentrosymmetric superconductor with in-plane magnetic fields.
Then, we confirm the validity of the edge theory by employing numerical simulations on a two-dimensional tight-binding model.
In Sec.~\ref{sec:3}, we study the effects of edge imperfections on the tunable MCMs.
Specifically, we calculate the local density of states (LDOS) for evaluating the stability and controllability of MCMs in non-square superconducting islands.
In Sec.~\ref{sec:4}, the differential conductance of normal lead wire/noncentrosymmetric superconductor hybrids is calculated by using lattice Green's function techniques.
We demonstrate that the zero-bias conductance shows the periodicity as a function of the direction of the applied magnetic field,
which serves as a fingerprint of the controllable MCMs.
Our conclusion is given in Sec.~\ref{sec:5}.

\section{Edge Theory and Tunable Majorana Corner Modes}
\label{sec:2}
\subsection{Model}
Let us consider a two-dimensional noncentrosymmetric ($s+p$)-wave superconductor in the presence of an in-plane magnetic field.
We start with a Bogoliubov-de Gennes (BdG) Hamiltonian in momentum space~\cite{sigrist_04,fujimoto_05,sigrist_06},
\begin{align}
\check{H}(\boldsymbol{k}) = \check{H}_{\mathrm{ncs}}(\boldsymbol{k}) + \check{H}_Z
\label{eq:bdg_bulk}
\end{align}
with
\begin{align}
&\check{H}_{\mathrm{ncs}}(\boldsymbol{k}) = \left[ \begin{array}{cc}
\hat{h}(\boldsymbol{k}) & \hat{\Delta}(\boldsymbol{k}) \\ -\hat{\Delta}^{\ast}(-\boldsymbol{k}) & -\hat{h}^{\ast}(-\boldsymbol{k}) \\ \end{array} \right],\\
&\hat{h}(\boldsymbol{k}) = \xi(\boldsymbol{k}) \hat{\sigma_0} + \lambda \boldsymbol{g}(\boldsymbol{k}) \cdot \hat{\boldsymbol{\sigma}},\\
&\hat{\Delta}(\boldsymbol{k}) = \left[ \Delta_s + \boldsymbol{d}(\boldsymbol{k}) \cdot \hat{\boldsymbol{\sigma}} \right]\left( i \hat{\sigma}_2 \right),\\
&\xi(\boldsymbol{k}) = \frac{\hbar^2 k^2}{2m} - \mu, \quad
\boldsymbol{d}(\boldsymbol{k}) = \Delta_t \frac{\boldsymbol{g}(\boldsymbol{k})}{k_F},
\end{align}
and
\begin{align}
\check{H}_Z = \left[ \begin{array}{cc}
\boldsymbol{V}_Z \cdot \hat{\boldsymbol{\sigma}} & 0 \\ 0 & -\left\{ \boldsymbol{V}_Z \cdot \hat{\boldsymbol{\sigma}} \right\}^{\ast} \\ \end{array} \right],
\end{align}
where $m$ is the effective mass of an electron, $\mu$ denotes the chemical potential,
$k_F = \sqrt{2m \mu}/\hbar$ represents the Fermi wave number, and $k=\sqrt{k^2_x + k^2_y}$.
The strength of Rashba spin-orbit coupling is given by $\lambda$ with $\boldsymbol{g}(\boldsymbol{k}) = (k_y, -k_x, 0)$.
The pair potential $\hat{\Delta}(\boldsymbol{k})$ contains both a spin-singlet $s$-wave component $\Delta_s$ and
a spin-triplet $p$-wave component $\boldsymbol{d}(\boldsymbol{k})$ satisfying
$\boldsymbol{d}(\boldsymbol{k}) \parallel \boldsymbol{g}(\boldsymbol{k})$~\cite{sigrist_04}.
The Zeeman potential induced by the externally applied in-plane magnetic field is described by
$\boldsymbol{V}_Z = (V_Z \cos \theta_Z, V_Z \sin \theta_Z, 0)$, with $\theta_Z$ representing the angle of the magnetic field measured from the $x$ direction.
In what follows, without loss of generality, we assume $\lambda$, $\Delta_s$, $\Delta_t$, $V_Z \geq 0$.
The Pauli matrices in spin space are given by $\hat{\boldsymbol{\sigma}} = (\hat{\sigma}_x, \hat{\sigma}_y, \hat{\sigma}_z)$,
and the $2 \times 2$ unit matrix is denoted by $\sigma_0$.

We briefly discuss the topological property of the pure noncentrosymmetric ($s+p$)-wave superconductor described by $\check{H}_{\mathrm{ncs}}(\boldsymbol{k})$.
The positive eigen energies of $\check{H}_{\mathrm{ncs}}(\boldsymbol{k})$ are given by
\begin{align} 
E^{\mathrm{ncs}}_{\eta} (\boldsymbol{k}) = \sqrt{
\left\{ \xi (\boldsymbol{k}) + \eta \lambda k \right\}^2 + \left\{ \Delta_t (k/k_F) + \eta \Delta_s \right\}^2 },
\end{align}
with $\eta=\pm$.
While the spectrum of $E^{\mathrm{ncs}}_+ (\boldsymbol{k})$ has a finite superconducting gap irrespective of the parameters,
the superconducting gap in $E^{\mathrm{ncs}}_- (\boldsymbol{k})$ vanishes when
\begin{align}
\Delta_s = \Delta_c,
\end{align}
with
\begin{align}
\Delta_c = \Delta_t \left[ \sqrt{1+ \left( \frac{\lambda k_F}{2 \mu} \right)^2 } + \frac{\lambda k_F}{2 \mu} \right].
\end{align}
For $\Delta_s > \Delta_c$ ($\Delta_s < \Delta_c$),
the BdG Hamiltonian $\check{H}_{\mathrm{ncs}}(\boldsymbol{k})$ can be deformed into
the BdG Hamiltonian of a pure spin-singlet $s$-wave (pure spin-triplet helical $p$-wave) superconductor without any gap closing
by decreasing $\lambda$ and $\Delta_t$ ($\Delta_s$) adiabatically.
The spin-triplet helical $p$-wave superconductor is well known as a time-reversal invariant topological superconductor characterized by a $\mathbb{Z}_2$ topological invariant
\cite{schnyder_08,zhang_08},
wheres the spin-singlet superconductor is topologically trivial.
Since the topological invariant never changes without gap closing,
the noncentrosymmetric ($s+p$)-wave superconductor with $\Delta_s < \Delta_c$ is topologically equivalent to the spin-triplet helical $p$-wave superconductor
and host the topologically protected helical Majorana edge states~\cite{sigrist_07,nagaosa_09,sato_09,sato_10,schnyder_15}.
In what follows, we only focus on the noncentrosymmetric ($s+p$)-wave superconductor in the topologically nontrivial phase (i.e., $\Delta_s < \Delta_c$).

Then, we discuss the effects of the Zeeman potential against the bulk energy spectrum.
Within first-order perturbation theory with respect to $V_Z$, the energy spectrum is given by
\begin{align}
E_{\eta} (\boldsymbol{k}) = E^{\mathrm{ncs}}_{\eta} (\boldsymbol{k}) + \eta V_Z \sin (\theta_{\boldsymbol{k}}-\theta_Z),
\end{align}
with $\theta_{\boldsymbol{k}}=\arctan (k_y/k_x)$.
Although the Zeeman potential causes a non-monotonic energy shift, it is clear that the Zeeman potential suppresses the superconducting gap size,
and a large enough Zeeman potential may bring the system into a gapless superconducting state
similar to the case of the pure helical $p$-wave superconductor discussed in Ref.~[\onlinecite{lee_13}].
Thus, throughout this paper, we only assume weak enough Zeeman potentials, such that the bulk superconducting gap remains finite and large enough.
Although the superconducting gap remains finite,
the noncentrosymmetric ($s+p$)-wave superconductor is no longer characterized by the $\mathbb{Z}_2$ topological invariant due to broken time-reversal symmetry.
Nevertheless, in the following, we demonstrate that
the noncentrosymmetric ($s+p$)-wave superconductor under the in-plane magnetic field becomes a second-order topological superconductor hosting MCMs.

\subsection{Effective Edge Hamiltonian}
\label{sec:2b}
\begin{figure}[tttt]
\begin{center}
\includegraphics[width=0.25\textwidth]{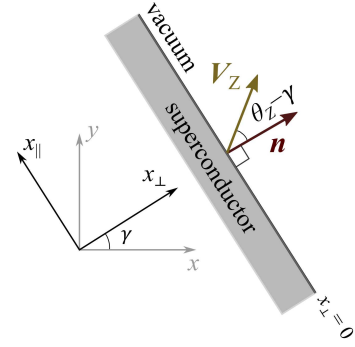}
\caption{Edge of the superconductor and relevant spatial coordinates.}
\label{fig:figure1}
\end{center}
\end{figure}

In order to discuss the emergence of MCSs in the present system intuitively, we derive an edge theory,
similar to the ones discussed in Refs.~[\onlinecite{wang_18, nori_18,zhu_18}].
We here assume that the present superconductor has an edge perpendicular to $\boldsymbol{n} = (\cos \gamma, \sin \gamma, 0)$,
where the spatial coordinate along the direction perpendicular (parallel) to the edge is represented by $x_{\perp}$ ($x_{\parallel}$),
and the superconductor occupies the entire half-space $x_{\perp}\leq0$ (see Fig.~\ref{fig:figure1}).
In addition, we apply periodic boundary conditions in the direction parallel to the edge (i.e., the direction along the $x_{\parallel}$-axis).
To obtain an effective edge Hamiltonian, we rewrite the BdG Hamiltonian in momentum space as,
\begin{align}
&\check{H}(\boldsymbol{k}) = \check{H}_0(\boldsymbol{k}) + \check{H}_{\parallel}(\boldsymbol{k}) + \check{H}_Z,\\
&\check{H}_0(\boldsymbol{k})=\left[ \begin{array}{cc}
\hat{h}_0(\boldsymbol{k}) & \hat{\Delta}_0 (\boldsymbol{k}) \\
-\hat{\Delta}_0^{\ast}(-\boldsymbol{k}) & -\hat{h}^{\ast}_0(-\boldsymbol{k}) \\ \end{array} \right],\\
&\hat{h}_0(\boldsymbol{k})=\xi(\boldsymbol{k}) \left(\frac{\hbar^2 k^2}{2m} - \mu\right) \hat{\sigma_0} + \lambda k_{\perp} \hat{\sigma}_{\parallel},\\
&\hat{\Delta}_0(\boldsymbol{k}) =\left[ \Delta_s +  \frac{\Delta_t}{k_F} k_{\perp} \hat{\sigma}_{\parallel} \right]\left( i \hat{\sigma}_2 \right) ,\\
&\check{H}_{\parallel}(\boldsymbol{k}) = \left[ \begin{array}{cc}
\hat{\lambda}_{\parallel}(k_{\parallel}) & \hat{\Delta}_{\parallel}(k_{\parallel}) \\
-\hat{\Delta}^{\ast}_{\parallel}(-k_{\parallel}) & -\hat{\lambda}^{\ast}_{\parallel}(-k_{\parallel}) \\ \end{array} \right],\\
&\hat{\lambda}_{\parallel}(k_{\parallel}) = \lambda k_{\parallel} \hat{\sigma}_{\perp}, \quad
\hat{\Delta}_{\parallel}(k_{\parallel}) = \frac{\Delta_t}{k_F} k_{\parallel}\hat{\sigma}_{\perp} \left( i \hat{\sigma}_2 \right),\\
&\check{H}_Z = \left[ \begin{array}{cc}
V_{\perp} \hat{\sigma}_{\perp}  + V_{\parallel} \hat{\sigma}_{\parallel} & 0
\\ 0 & -\left\{ V_{\perp} \hat{\sigma}_{\perp} + V_{\parallel} \hat{\sigma}_{\parallel} \right\}^{\ast} \\ \end{array} \right],
\end{align}
with
\begin{align}
&k_{\perp} = k_x \cos \gamma + k_y \sin \gamma, \\
&k_{\parallel} = -k_x \sin \gamma + k_y \cos \gamma, \\
&\hat{\sigma}_{\perp} = \hat{\sigma}_x \cos \gamma + \hat{\sigma}_y \sin \gamma, \\
&\hat{\sigma}_{\parallel} = -\hat{\sigma}_x \sin \gamma + \hat{\sigma}_y \cos \gamma, \\
&V_{\perp} = V_Z \cos(\theta_Z - \gamma), \quad V_{\parallel} = V_Z \sin(\theta_Z - \gamma),
\end{align}
where $k_{\perp}$($k_{\parallel}$) represent the momentum perpendicular (parallel) to the edge of the superconductor.
In what follows, we treat $\check{H}_{\parallel}(\boldsymbol{k})$ and $\check{H}_Z$ as the perturbations.
This approximation is justified when $k_{\parallel}/k_F \ll 1$ and when $V_Z$ is significantly smaller than the superconducting gap.
Moreover, in the following analysis, we assume $\Delta_s, \Delta_t \ll \lambda k_F, \mu$. 

For the zeroth-order perturbation, we replace $k_{\perp} \rightarrow - i\partial_{x_\perp}$ and find zero-energy states satisfying the equation,
\begin{align}
\check{H}_0(- i\partial_{x_\perp}, k_{\parallel}) \psi(x_{\perp}) = 0,
\end{align}
with the boundary condition $\psi(0) = \psi(-\infty) = 0$.
As a result, we find the two solutions,
\begin{align}
&\psi_{+, k_{\parallel}}(x_{\perp}) = \frac{1}{2} \left[ \begin{array}{c}
e^{-i \gamma/2} \\ i e^{i \gamma/2} \\ - e^{i \gamma/2} \\ -i e^{- i \gamma/2} \end{array} \right] \varphi_+(x_{\perp}),\\
&\psi_{-, k_{\parallel}}(x_{\perp}) = \frac{1}{2} \left[ \begin{array}{c}
- i e^{-i \gamma/2} \\ - e^{i \gamma/2} \\ i e^{i \gamma/2} \\ e^{- i \gamma/2} \end{array} \right] \varphi_-(x_{\perp}),\\
&\varphi_{\zeta}(x_{\perp}) = 2 \sqrt{\kappa} \sin (\sqrt{k^2_F + k^{2}_{\lambda}} x_{\perp}) e^{i \zeta k_{\lambda} x_{\perp}} e^{ \kappa x_{\perp}},\\
&\kappa = \frac{m \Delta_t}{\hbar^2 k_F}, \quad k_{\lambda} = \frac{m \lambda}{\hbar^2}
\end{align}
with $\zeta = \pm$, where we neglect the insignificant terms of order $O(\frac{k^2_{\parallel}}{k^2_F}, \frac{\Delta_s}{\mu}, \frac{\Delta_t}{\mu})$. 
The matrix elements of the perturbation terms $\check{H}_{\parallel}(\boldsymbol{k}) + \check{H}_Z$ within the first order are calculated as
\begin{align}
a_{\zeta,\zeta^{\prime}} = \int^{0}_{-\infty} d x_{\perp}
\psi^{\dagger}_{\zeta}(x_{\perp}) \left\{ \check{H}_{\parallel}(\boldsymbol{k}) + \check{H}_Z \right\} \psi_{\zeta^{\prime}} (x_{\perp}).
\end{align}
As a result, we obtain an effective edge Hamiltonian,
\begin{align}
&{\cal H}^{\prime} = \left[ \begin{array}{cc} a_{+,+} & a_{+,-} \\ a_{-,+} & a_{-,-} \\ \end{array} \right]
= \left[ \begin{array}{cc}
V_{\parallel} & \alpha k_{\parallel} \\
\alpha^{\ast} k_{\parallel} & -V_{\parallel} \\ \end{array} \right],\\
& \alpha = \frac{\beta (\Delta_t - i \lambda k_F)}{k_F}, \quad
\beta = \frac{\Delta_t}{\lambda k_F} \frac{4 \mu^2 + \left( \lambda k_F \right)^2 }{4 \mu^2},
\end{align}
which is unitarily equivalent to
\begin{align}
{\cal H} = \left[ \begin{array}{cc}
\gamma k_{\parallel} &  V_{\parallel} \\  V_{\parallel} & - \gamma k_{\parallel} \\ \end{array} \right],
\label{eq:edge_ham}
\end{align}
with $\gamma = \beta \sqrt{\Delta^2_t + \left(\lambda k_F\right)^2 } /k_F$.
From Eq.~(\ref{eq:edge_ham}), we clearly find that the mass term for the linearly dispersive helical edge states is given by $V_{\parallel}=V_Z \sin(\theta_Z - \gamma)$.
Importantly, the sign of the mass term is determined only by the relative angle between the Zeeman field $\boldsymbol{V}_Z$ and the edge normal vector $\boldsymbol{n}$:
\begin{align}
\begin{split}
& V_{\parallel} > 0  \quad \text{for} \quad 0 < \theta_Z - \gamma < \pi,\\
& V_{\parallel} < 0  \quad \text{for} \quad -\pi < \theta_Z - \gamma < 0,\\
& V_{\parallel} = 0  \quad \text{for} \quad \theta_Z - \gamma = 0,\pi. \label{eq:mass-term}
\end{split}
\end{align}
Thus, there is a possibility that two adjacent edges of the system have the mass terms with opposite signs.
In such case, we obtain zero energy states bounded in the vicinity of the corner (i.e., MCMs) because the mass term has a kink there~\cite{wang_18, nori_18,zhu_18}.
In the next section, we confirm the validity of our edge theory by performing numerical calculations of a tight-binding model.

\subsection{Tunable Majorana corner modes}
\label{sec:2c}
\begin{figure}[tttt]
\begin{center}
\includegraphics[width=0.425\textwidth]{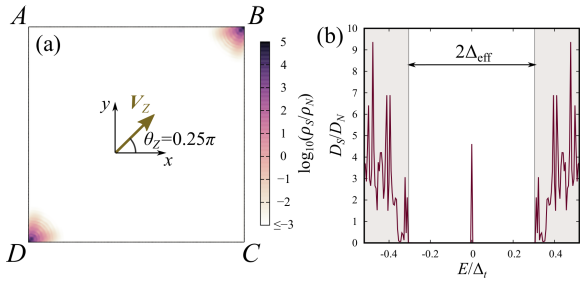}
\caption{(a) LDOS at zero energy of the tight-binding model on a square lattice with $(201 \times 201)$ lattice sites.
Specifically, we plot the spatial distribution of $\log_{10} (\rho_S/\rho_N)$.
The four corners are labeled as $A$, $B$, $C$, and $D$, respectively.
(b) DOS as a function of energy $E$.
}
\label{fig:figure2}
\end{center}
\end{figure}
We here consider a square superconducting island as shown in Fig.~\ref{fig:figure2}(a), where the four corners are labeled as $A$, $B$, $C$, and $D$, respectively.
When we apply the Zeeman field along $\theta_Z=0.25\pi$, according to Eq.~(\ref{eq:mass-term}),
the edges $AB$ and $DA$ have negative mass terms, whereas the edges $BC$ and $CD$ have positive mass terms.
Therefore, based on the effective edge theory, we can expect MCMs at the corners $B$ and $D$.
To confirm the above statement, we numerically calculated the LDOS by using the formula
$\rho_S(\boldsymbol{r},E) = - \mathrm{Tr}\left[ \mathrm{Im} \left\{ \check{G}(\boldsymbol{r},\boldsymbol{r},E+i\delta) \right\}\right]/\pi$,
where $\check{G}(\boldsymbol{r},\boldsymbol{r}^{\prime},E)$ represents the Green's function
and $\mathrm{Tr}$ denotes the trace in spin and Nambu spaces.
$\delta$ is a small imaginary part added to the energy $E$.
The Green's function is calculated on a tight-binding model by use of a recursive Green's function technique~\cite{fisher_81}, where we replace
$\xi(\boldsymbol{k}) \rightarrow 2t( 1- \cos k_x) + 2t( 1- \cos k_y) -\mu$,
$\boldsymbol{g}(\boldsymbol{k}) \rightarrow (\sin k_y, -\sin k_x, 0)$, and $\Delta_t/k_F \rightarrow \Delta_t$.
The explicit expression of the BdG tight-binding Hamiltonian is given in Appendix~\ref{sec:appA}.
In Fig.~\ref{fig:figure2}(a), we show the LDOS at zero-energy for a $(201 \times 201)$ square island.
We use the parameters: $\mu=t$, $\lambda=0.5t$, $\Delta_t=0.5t$, $\Delta_s=0.2\Delta_t$, $V_Z = 0.5\Delta_t$, and $\delta=10^{-4}\Delta_t$.
The LDOS is normalized by $\rho_N = \langle \rho_N(\boldsymbol{r},E=0) \rangle$, where
$\rho_N(\boldsymbol{r},E)$ representing the LDOS for the normal state (i.e., $\Delta_t=\Delta_s=0$),
and $\langle  \cdots \rangle$ means the averaged value with respect to the lattice sites.
Moreover, we plot $\log_{10} (\rho_S/\rho_N)$ instead of the raw data of $\rho_S$.
As shown in Fig.~\ref{fig:figure2}(a), the LDOS at zero energy has a significant amplitude only in the vicinity of the the corners $B$ and $D$.
This result confirms the presence of MCMs, as expected. 
In Fig.~\ref{fig:figure2}(b), we show the density of states (DOS) as a function of the energy.
The DOS is calculated by $D_S(E) = \sum_{\boldsymbol{r} \in S} \rho_S (\boldsymbol{r},E)$,
where $\sum_{\boldsymbol{r} \in S}$ represents the summation over the lattice sites in the superconductor island.
The results are normalized by the DOS in the normal state, $D_N = (201 \times 201) \rho_N$.
As shown in Fig.~\ref{fig:figure2}(b), the DOS exhibits a sharp zero energy peak inside the superconducting gap $\Delta_{\mathrm{eff}} = 0.31\Delta_t$.
Due to particle-hole symmetry of the superconductor,
the two MCMs can depart from zero energy only when they hybridize with each other, thereby becoming conventional electron-like and hole-like quasiparticle modes. 
Therefore, as long as the two MCMs are spatially separated and protected by the superconducting gap $\Delta_{\mathrm{eff}}$, they can remain robust at zero energy.

\begin{figure}[bbbb]
\begin{center}
\includegraphics[width=0.4\textwidth]{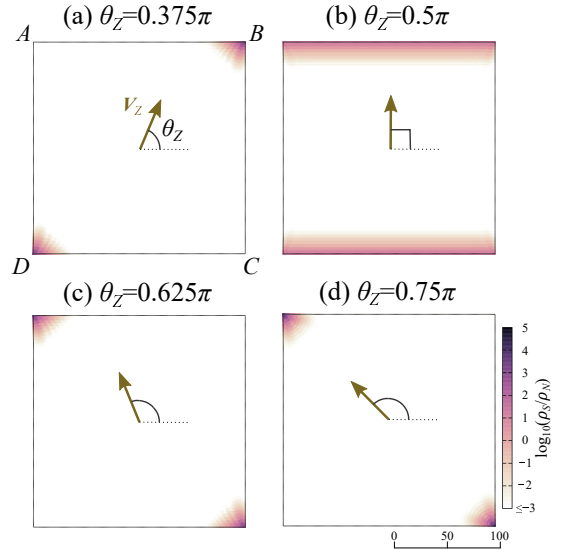}
\caption{LDOS at zero energy with (a) $\theta_Z= 3\pi/8$, (b) $\pi/2$, (c) $5\pi/8$, and (d) $3\pi/4$.}
\label{fig:figure3}
\end{center}
\end{figure}
Next, we discuss the controllability of the MCMs.
In Fig.~\ref{fig:figure3}, we show the LDOS at zero energy with various magnetic field directions,
namely (a) $\theta_Z= 0.375\pi$, (b) $0.5\pi$, (c) $0.625\pi/8$, and (d) $0.75\pi$.
When $\theta_Z= 0.375\pi$, according to Eq.~(\ref{eq:mass-term}), the kinks of the mass term are located at the corners $B$ and $D$.
As a result, although the spatial distributions of the MCMs are slightly modulated from the case with $\theta_Z=0.25\pi$,
we can still find them at the corners $B$ and $D$.
When $\theta_Z= 0.5\pi$, the mass term becomes zero at the entire edges $AB$ and $CD$.
As a consequence, zero-energy Majorana {\it edge} states appear on the entire edges $AB$ and $CD$.
The emergence of Majorana edge states in a time-reversal invariant topological superconductor under a Zeeman field with a certain direction
has been discussed in terms of the Majorana Ising spin~\cite{sato_09,zhang_09,nagaosa_10,yamakage_17},
and it has been shown that the energy spectrum of the edge states becomes gapless
(strictly speaking, the energy spectrum is discrete due to the finite size effect)~\cite{wang_18,sato_09,zhang_09,nagaosa_10,yamakage_17}.
When $\theta_Z$ exceeds $0.5\pi$, the positions of the mass term kink move to the corner $A$ and $C$.
As a consequence, for $\theta_Z= 0.625\pi$ and $0.75\pi$, the MCMs appear at the corners $A$ and $C$.
By varying the direction of the Zeeman field from $\theta_Z=0.25\pi$ to $0.75\pi$, the MCM originally located at the corner B hop to the corner A.
In the same way, the pair of the MCMs can make a full circle around the system by rotating the direction of the Zeeman field by $2\pi$.

\section{Effect of edge imperfections}
\label{sec:3}
In the above, we have demonstrated that the noncentrosymmetric ($s+p$)-wave superconductor can host tunable MCMs.
However, in real experiments, the superconducting island generally deviates form the perfect square shape.
Thus, in this section we study the effects of different island shapes (subsection \ref{sec:3a}) and of edge disorder (subsection \ref{sec:3b}).

\subsection{Non-square geometry}
\label{sec:3a}
\begin{figure}[hhhh]
\begin{center}
\includegraphics[width=0.225\textwidth]{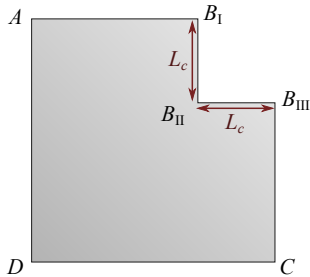}
\caption{Non-square island obtained by removing an ($L_c \times L_c$) square from the upper right corner of a bigger square.}
\label{fig:figure4}
\end{center}
\end{figure}
\begin{figure}[tttt]
\begin{center}
\includegraphics[width=0.4\textwidth]{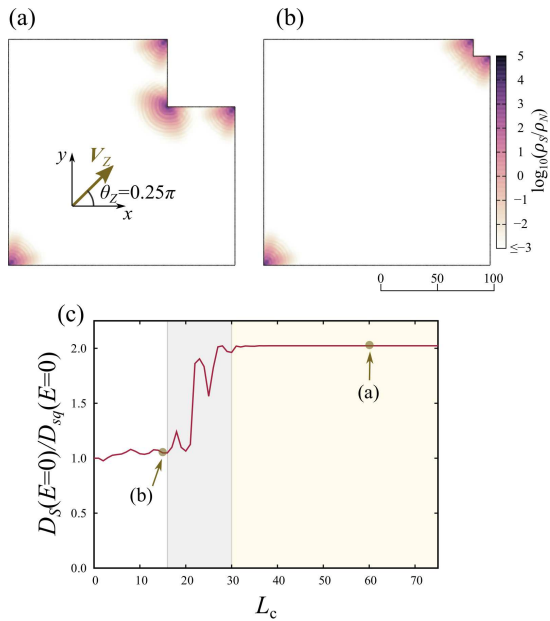}
\caption{LDOS at zero energy for the non-square island of Fig.~\ref{fig:figure4} with (a) $L_c =60$ and (b) $L_c =15$.
In (c), we show the DOS at zero energy as a function of $L_c$,
where the DOS is normalized by $D_{\mathrm{sq}}(E=0)$ representing the DOS of the ideal square island.}
\label{fig:figure5}
\end{center}
\end{figure}

As an example of a non-square shape, we consider a square island,
where a square regions of size ($L_c \times L_c$) has been removed from the upper right corner $B$~\cite{wang_18}, as shown in Fig.~\ref{fig:figure4}.
We label the six corners as $A$, $B_{\mathrm{I}}$, $B_{\mathrm{II}}$, $B_{\mathrm{III}}$, $C$, and $D$.
We first consider the MCMs for a field applied along the direction $\theta_Z = 0.25 \pi$.
Under this circumstance, according to Eq.~(\ref{eq:mass-term}),
the mass term changes its sign at the four corners $B_{\mathrm{I}}$, $B_{\mathrm{II}}$, $B_{\mathrm{III}}$, and $D$.
In Fig.~\ref{fig:figure5}, we show the LDOS at zero energy for an island with the shape of Fig.~\ref{fig:figure4} with (a) $L_c =60$ and (b) $L_c =15$.
In accordance with the edge theory, as shown in Fig.~\ref{fig:figure5}(a), we find four MCMs for the island with $L_c = 60$.
However, for an island with $L_c = 15$,
there are only two MCMs, where one of them is spread across the corners $B_{\mathrm{I}}$, $B_{\mathrm{II}}$, and $B_{\mathrm{III}}$.
This implies that the three MCMs originally located at $B_{\mathrm{I}}$, $B_{\mathrm{II}}$ and $B_{\mathrm{III}}$ hybridize with each other
thereby becoming non-degenerate, with two states at small non-zero energy and one state at zero energy, in accordance with particle-hole symmetry. 
In fact, the edge theory is valid only when the edges are sufficiently longer than the decay length of the MCMs
because, to derive the edge theory, we apply the periodic boundary condition in the direction parallel to the edge and do not take into account the length of the edge.
Roughly speaking, the decay length of the MCM is evaluated by the inverse of the superconducting gap~\cite{lee_10},
i.e., $\xi_{\mathrm{eff}} \approx t/\Delta_{\mathrm{eff}}$, where $\xi_{\mathrm{eff}}=6.5$ with the present parameter choices.
Thus, when $L_c$ exceeds $\xi_{\mathrm{eff}}$ significantly, we obtain the MCMs at the three corners $B_{\mathrm{I}}$, $B_{\mathrm{II}}$ and $B_{\mathrm{III}}$.  
In Fig.~\ref{fig:figure5}(c), we show the DOS at zero energy as a function of $L_c$,
where the DOS is normalized by $D_{\mathrm{sq}}(E=0)$ representing the zero-energy DOS for the ideal square island, calculated in Fig.~\ref{fig:figure2}(b).
When $L_c$ is large enough (i.e., $L_c \gtrsim 30$), we find $D_S/D_{\mathrm{sq}}=2$.
Although the DOS does not count directly the number of the states, the relation of $D_S/D_{\mathrm{sq}}=2$ implies that
the number of the MCMs in the present island is twice as much as that in the ideal square island: there are four MCMs as shown in Fig.~\ref{fig:figure4}(a).
For $L_c \lesssim 16$, we find $D_S/D_{\mathrm{sq}} \approx 1$ suggesting the presence of
two MCMs as shown in Fig.~\ref{fig:figure4}(b).
In the intermediate region (i.e. $16 \lesssim L_c \lesssim 30$), the DOS shows non-monotonic oscillations,
which originate from interference effects between the three MCMs, which modify their mutual couplings.

\begin{figure}[tttt]
\begin{center}
\includegraphics[width=0.475\textwidth]{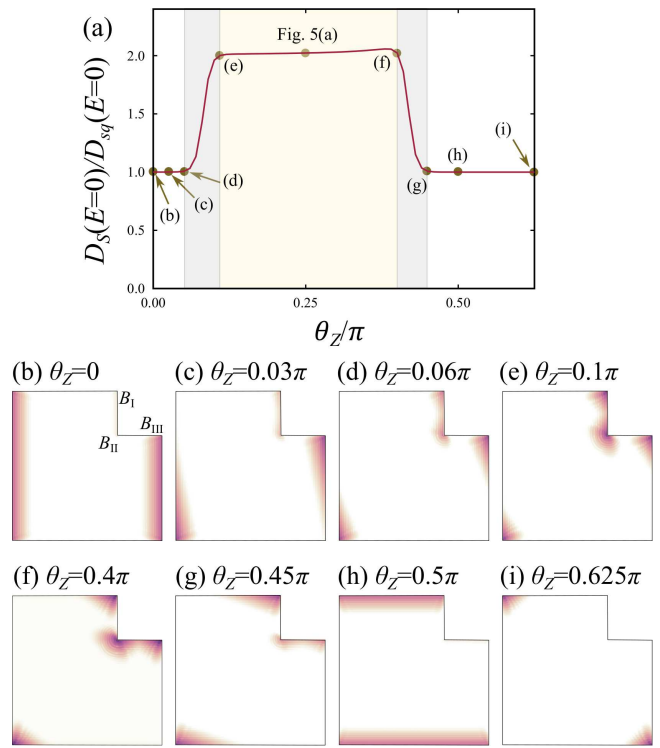}
\caption{(a) DOS at zero energy as a function of the angle of the Zeeman field $\theta_Z$, where we choose $L_c=60$.
(b)-(i) LDOS at zero energy for the eight different values of $\theta_Z$ indicated in (a). }
\label{fig:figure6}
\end{center}
\end{figure}
In Fig.~\ref{fig:figure6}(a), we show the DOS at zero energy as a function of the angle of the Zeeman field,
where we vary $0 \leq \theta_Z \leq 0.625\pi$ and choose $L_c=60$. 
We find $D_S/D_{\mathrm{sq}} \approx 2$ for $0.1 \pi \lesssim \theta_Z \lesssim 0.4\pi$,
whereas $D_S/D_{\mathrm{sq}} \approx 1$ for $0 \lesssim \theta_Z \lesssim 0.06\pi$ and $0.45 \pi \lesssim \theta_Z$.
Therefore, the number of the MCMs changes by rotating the Zeeman potential.
To see more details, in Figs.~\ref{fig:figure6}(b)-\ref{fig:figure6}(i), we present the LDOS at zero energy for eight different values of $\theta_Z$.
When $\theta_Z=0$, as shown in Fig.~\ref{fig:figure6}(b), we find the Majorana edge states at the edge $B_{\mathrm{III}}C$,
whereas there are no significant enhancement in the LDOS at the edge $B_{\mathrm{I}}B_{\mathrm{II}}$.
When $\theta_Z=0.03\pi$, as shown in Fig.~\ref{fig:figure6}(c), an MCM appears at the corner $B_{\mathrm{III}}$, where the wave function is extended towards the corner $C$.
In addition, we find the small enhancement of the LDOS at the edge $B_{\mathrm{I}}B_{\mathrm{II}}$.
Strictly speaking, this small enhancement is not originated from exact zero energy states
but from the finite energy states having energy extremely close to zero. 
We cannot eliminate contributions from such low-energy states because, although the strict definition of the LDOS is given by
$\rho_S(\boldsymbol{r},E) = - \lim_{\delta \to 0+} \mathrm{Tr}\left[ \mathrm{Im} \left\{ \check{G}(\boldsymbol{r},\boldsymbol{r},E+i\delta) \right\}\right]/\pi$,
the small imaginary part of $\delta$ is always set to be finite in our numerical calculations.
As shown in Fig.~\ref{fig:figure6}(d), the LDOS at the edge $B_{\mathrm{I}}B_{\mathrm{II}}$ increases by increasing $\theta_Z$.
When $\theta_Z=0.1\pi$, as shown in Fig.~\ref{fig:figure6}(e),
we find that the zero-energy LDOS is significantly enhanced in the vicinity of the {\it corners} $B_{\mathrm{I}}$ and $B_{\mathrm{II}}$.
As already shown in Fig.~\ref{fig:figure5}(a), with $\theta_Z=0.25\pi$, we obtain three distinct MCMs at the three corners
$B_{\mathrm{I}}$, $B_{\mathrm{II}}$, and $B_{\mathrm{III}}$.
As a consequence, by increasing $\theta_Z$ from zero to $0.25\pi$,
two additional MCMs are created at the corners $B_{\mathrm{I}}$ and $B_{\mathrm{II}}$, while a single MCM stays constantly at the corner $B_{\mathrm{II}}$. 
When $\theta_Z=0.4\pi$, as shown in Fig.~\ref{fig:figure6}(e),
the wave function of the MCM at the corner $B_{\mathrm{III}}$ is extended towards the corner $B_{\mathrm{II}}$,
such that the MCMs at the corners $B_{\mathrm{II}}$ and $B_{\mathrm{III}}$ start to overlap with each other. 
When $\theta_Z=0.45\pi$, as shown in Fig.~\ref{fig:figure6}(e), the LDOS at the edge $B_{\mathrm{II}}B_{\mathrm{III}}$ is strongly suppressed.
With $\theta_Z=0.5\pi$, we can no longer find the enhancement of the LDOS at the edge $B_{\mathrm{II}}B_{\mathrm{III}}$,
whereas we find a single Majorana edge state at the edge $AB_{\mathrm{I}}$.
With $\theta_Z=0.625\pi$, we find only two MCMs at the corner $A$ and $C$.
Therefore, by increasing $\theta_Z$ from $0.25\pi$ to $0.625 \pi$,
the MCMs at the corner $B_{\mathrm{II}}$ and that at $B_{\mathrm{III}}$ start to hybridize with each other and move away from zero energy.
At the same time the single MCM originally located at the corner $B_{\mathrm{I}}$ hops to the corner $A$, while the single MCM of corner $D$ hops to corner $C$.
Importantly, in this process,
the MCMs originally appearing at $B_{\mathrm{II}}$ and $B_{\mathrm{III}}$ vanish by hybridizing with each other, and never reappear at any other corner.
To accomplish the braiding process in topological quantum computations, we must exchange the positions of the two Majorana zero modes forming a pair.
In the present edge configuration, however,
we can no longer find a stable pair of MCMs that can fully circle the edges of the superconductor, and therefore fail the braiding process.
To avoid the emergence of additional and undesired MCMs, as also shown in Fig.~\ref{fig:figure5},
we have to eliminate additional corners whose adjacent edges are longer that the decay length of the MCMs.

\subsection{Edge roughness}
\label{sec:3b}
\begin{figure}[bbbb]
\begin{center}
\includegraphics[width=0.45\textwidth]{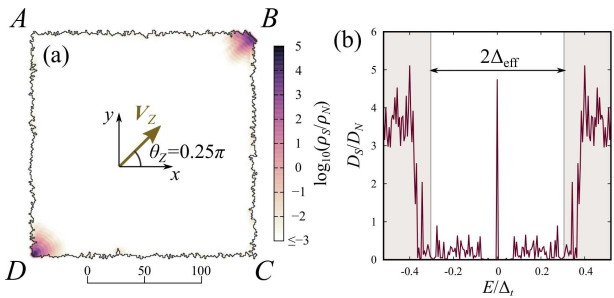}
\caption{(a) LDOS at zero energy and (b) DOS as a function of the energy $E$ for the superconducting island with the rough edge $(p,X)=(0.1,20)$.}
\label{fig:figure7}
\end{center}
\end{figure}
Next, we consider a square island with rough edges.
To described the rough edges, we remove the outermost sites from an ideal square island with probability $p$,
and repeat this etching process $X$ times~\cite{guinea_10,schnyder_14}.
With this process, there is a possibility that small debris separated from the largest island are created. Thus, we also remove such debris.
In Figs.~\ref{fig:figure7}(a) and \ref{fig:figure7}(b), we show the LDOS at zero energy and the DOS as a function of the energy, respectively.
We apply the Zeeman field along $\theta_Z=0.25\pi$ and consider a rough edge with $(p,X)=(0.1,20)$.
We find that the zero-energy LDOS in the vicinity of the corners $B$ and $D$ is enhanced significantly.
In addition, as shown in Fig.~\ref{fig:figure7}(b), the DOS retains a steep zero energy peak structure,
whereas we also find finite DOS inside the effective superconducting gap in the clean limit $\Delta_{\mathrm{eff}} = 0.31\Delta_t$.
As also discussed in Sec.~\ref{sec:2c}, the MCMs are protected by the particle-hole symmetry as long as they are sufficiently separated spatially.
Thus, the MCMs remain robust even in the presence of weak or moderate edge roughness that does not destroy the superconducting gap completely.

\begin{figure}[tttt]
\begin{center}
\includegraphics[width=0.375\textwidth]{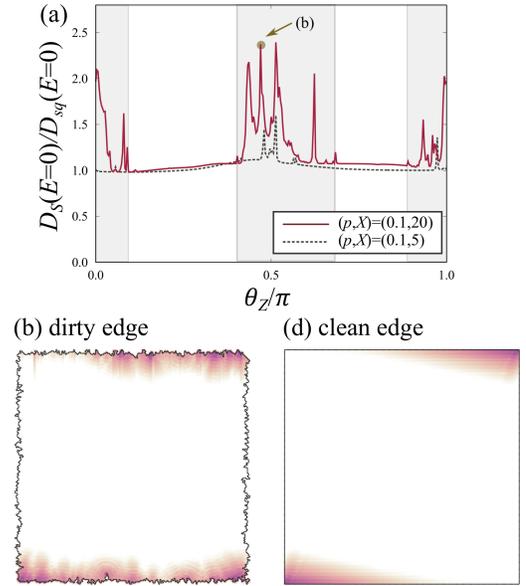}
\caption{(a) DOS at zero energy as a function of the angle of the Zeeman field $\theta_Z$.
The solid-red line denotes the result with the same island as in Fig.~\ref{fig:figure7}(a) [i.e., $(p,X)=(0.1,20)$],
and the dotted black line denotes the result for the island with weaker roughness $(p,X)=(0.1,5)$.
(b)-(c) LDOS at zero energy with $\theta_Z = 0.47\pi$.
In (b), we use the same island as in Fig.~\ref{fig:figure7}(a) [i.e., $(p,X)=(0.1,20)$].
In (c), for comparison, we show the result with clean edges.}
\label{fig:figure8}
\end{center}
\end{figure}
In Fig.~\ref{fig:figure8}(a), we show the normalize DOS at zero energy as a function of the direction of the Zeeman field.
For the solid red line, we use the same superconducting island as in Fig.~\ref{fig:figure7}(a), i.e., $(p,X)=(0.1,20)$.
For the dotted black line, we consider an island with weaker roughness $(p,X)=(0.1,5)$.
At first, we concentrate on the results with $(p,X)=(0.1,20)$, shown by the solid-red line.
For $0.1\pi \lesssim \theta_Z \lesssim 0.4 \pi$ and for $0.69\pi \lesssim \theta_Z \lesssim 0.89 \pi$,
we find $D_S/D_{\mathrm{sq}} \approx 1$ suggesting the presence of distinct MCMs.
However, in the vicinity of $\theta_Z=0$, $0.5\pi$, and $\pi$, the DOS shows a non-monotonic dependence and
becomes clearly larger than one, i.e., $D_S/D_{\mathrm{sq}} > 1$.
To see more details, in Fig.~\ref{fig:figure8}(b), we show the LDOS at zero-energy with $\theta_Z = 0.47\pi$,
where we use the same superconducting island for calculating the solid-red line in Fig.~\ref{fig:figure8}(a).
For comparison, in Fig.~\ref{fig:figure8}(c), we also show the result with the clean edge.
With the dirty edge, we find that the LDOS is enhanced not only in the vicinity of the corner $B$ and $D$
but also around the center of the edge $AB$ and at the corner $C$. 
According to the effective edge theory, when $\theta_Z$ is close to $0.5 \pi$,
the mass term $V_{\parallel}=V_Z \sin(\theta_Z - \gamma)$, which is the source of the energy gap in the edge spectrum, is suppressed at the edges $AB$ and $CD$.
Particularly, the mass term of the edges $AB$ and $CD$ vanishes completely at $\theta = 0.5 \pi$.
The results in Fig.~\ref{fig:figure8} suggest that
the already small energy gap is reduced further by the edge roughness, such that states with extremely small (but finite) energy appear at the edges $AB$ and $CD$.
In the topological quantum computations, the energy gap which energetically separates the MCMs from other finite energy states
plays an essential role to suppress the decoherence during the braiding operations.
Therefore, the edge roughness causing damage to the energy gap in the edge spectrum affects negatively the braiding processes of the MCMs.
As shown by the dotted black line in Fig.~\ref{fig:figure8}(a), the undesirable enhancement in the low-energy DOS is suppressed by decreasing the edge roughness.

In the above, we have shown that, although the MCMs exist robustly, the braiding operation of them may be disturbed by edge imperfections.
The simplest countermeasure would be fabricating superconducting islands with clean and smooth edges.
Even so, establishing a practical way to to make the braiding process more robust would be a desirable work~\cite{trauzettel_20a,soluyanov_20,trauzettel_20}.

\section{Tunneling spectroscopy}
\label{sec:4}
\begin{figure}[bbbb]
\begin{center}
\includegraphics[width=0.35\textwidth]{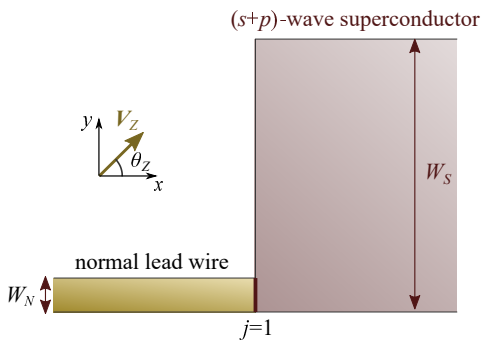}
\caption{Schematic image of the normal lead wire/($s+p$)-wave superconductor junction.}
\label{fig:figure9}
\end{center}
\end{figure}
Finally, we propose an experiment for detecting the tunable MCMs.
Let us consider a normal lead wire/superconductor junction, where the normal lead wire is attached to a corner of the superconductor as shown in Fig.~\ref{fig:figure9}.
We describe the present junction with a tight-binding model, where a lattice site is indicated by $\boldsymbol{r}=j \boldsymbol{x}+ m \boldsymbol{y}$
with $\boldsymbol{x}$ ($\boldsymbol{y}$) representing the unit vector in the $x$ ($y$) direction.
The superconductor (lead wire) is located at $ 1 \leq j \leq \infty$ ($-\infty \leq j < 1$) and $1 \leq m \leq W_S$ ($1 \leq m \leq W_N$),
where $W_S$ ($W_N$) denotes the width of the superconductor (lead wire).
The superconducting segment is described by the BdG Hamiltonian used also in the above numerical calculations,
and the normal lead wire is described by setting $\Delta_t=\Delta_s=0$ and $\lambda=0$.
The hopping integral between the superconductor and the lead wire (i.e., $j=0$ and $j=1$) is given by $t_{\mathrm{int}}$.
The explicit form of the BdG Hamiltonian describing the present junction is given in Appendix~\ref{sec:appA}.
In the following calculations, we fix the parameters as
$\mu=t$, $\lambda=0.5t$, $\Delta_t=0.5t$, $\Delta_s=0.2\Delta_t$, $V_Z = 0.5\Delta_t$, $t_{\mathrm{int}}=0.1t$, $W_N=10$, and $W_S=101$.
We assume a sufficiently low transparency at the junction interface (i.e., $t_{\mathrm{int}}=0.1t$) such that the bias voltage is mainly dropped at the interface.
Under this circumstance, we can calculate the differential conductance at zero temperature by using the Blonder--Tinkham--Klapwijk formula
\cite{klapwijk_82, bruder_90, kashiwaya_00},
\begin{align}
G(eV) = \frac{e^{2}}{h} \sum_{\zeta,\zeta^{\prime}}
\left[ \delta_{\zeta,\zeta^{\prime}} - \left| r^{ee}_{\zeta,\zeta^{\prime}} \right|^{2}
+ \left| r^{he}_{\zeta,\zeta^{\prime}} \right|^{2} \right]_{E=eV},
\end{align}
where $r^{ee}_{\zeta,\zeta^{\prime}}$ and $r^{he}_{\zeta,\zeta^{\prime}}$ denote a normal and Andreev reflection coefficient at energy $E$, respectively.
The indexes $\zeta$ and $\zeta^{\prime}$ label an outgoing and incoming channel in the normal lead wire, respectively.
These reflection coefficients are calculated using recursive Green's function techniques~\cite{fisher_81, ando_91}.

\begin{figure}[tttt]
\begin{center}
\includegraphics[width=0.48\textwidth]{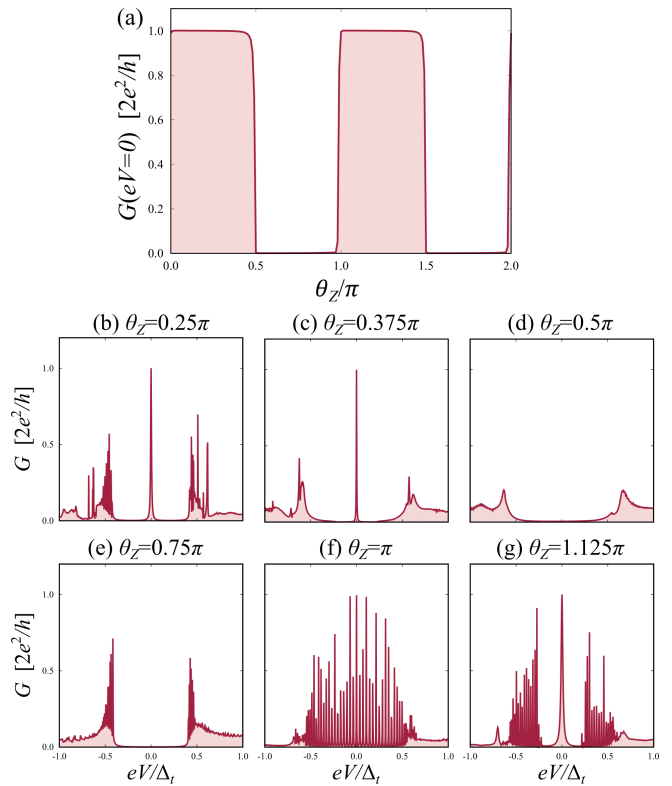}
\caption{(a) Zero-bias conductance as a function of the angle of the Zeeman field $\theta_Z$.
(b)-(g) Differential conductance as a function of the bias voltages for (b) $\theta_Z=0.25\pi$, (c) $0.375\pi$, (d) $0.5\pi$, (e) $0.75\pi$, (f) $\pi$, and (f) $1.125\pi$.}
\label{fig:figure10}
\end{center}
\end{figure}

In Fig.~\ref{fig:figure10}(a), we show the zero-bias conductance $G(eV=0)$ as a function of the angle of the Zeeman field $\theta_Z$.
We find the conductance plateaus at $2e^2/h$ for $0<\theta_Z<0.5\pi$ and $\pi<\theta_Z<1.5\pi$.
For these $\theta_Z$, according to the effective edge theory in Sec.~\ref{sec:2b}, the MCM appears at the corner attached to the lead wire.
Therefore, it is clear that the conductance plateaus is caused by the resonant tunneling through the MCM.
To see more details, in Fig.~\ref{fig:figure10}(b)-\ref{fig:figure10}(g), we show the differential conductance as a function of the bias voltage for various $\theta_Z$.
At $\theta_Z=0.25\pi$, as shown in Fig.~\ref{fig:figure10}(b), we find a steep zero-bias conductance peak,
where the conductance spectrum for finite energies displays an almost hard-gap structure.
For $\theta_Z=0.375\pi$, we still find the clear peak, whereas the superconducting gap edge has now a tail that was absent for $\theta_Z=0.25\pi$.
When $\theta_Z=0.5\pi$, the Majorana {\it edge} mode appears along the edge perpendicular to the junction interface.
Since the Majorana edge mode is not bounded along the junction interface, it cannot cause resonant tunneling~\cite{asano_04}.
As a result, as shown in Fig.~\ref{fig:figure10}(d), the zero-bias conductance peak disappears when $\theta_Z=0.5\pi$.
At $\theta_Z=0.75\pi$, there are no zero-energy states in the vicinity of the corner attached to the lead wire.
As a consequence, we find a hard gap in the conductance spectrum.
When $\theta_Z=\pi$, the gapless Majorana edge states appear at the edge along the junction interface.
Here, we must note that the energy levels of the Majorana edge states are discretized owing to the finite size of the system.
As a consequence, as shown in Fig.~\ref{fig:figure10}(f), the conductance spectrum exhibits many spikes for low bias voltages.
When we prepare a sufficiently large superconductor, the energy levels of the Majorana edge states will form a continuum,
such that the conductance spectra show a broad peak structure instead of a collection of sharp peaks.
With $\theta_Z=1.125\pi$, the MCM again appears at the corner attached to the lead wire.
As a consequence, as shown in Fig.~\ref{fig:figure10}(g), the conductance spectrum exhibits a zero bias peak,
whereas we also find significant enhancement in low bias voltages.
When $\theta_Z=1.25\pi$, the conductance spectrum returns to that in Fig.~\ref{fig:figure10}(b).

The periodicity of the zero-bias conductance with respect to the direction of the Zeeman field is a smoking gun evidence for the existence of controllable MCMs.
In real experiments, finite temperature effects and other perturbations may modify the details of the conductance spectrum.
Moreover, the edge roughness may disturb the stepwise change in the zero-bias conductance shown in Fig.~\ref{fig:figure10}(a)
because according to Sec.~\ref{sec:3b}, the edge roughness affects significantly the spectrum of the edge states around $\theta=0$, $0.5\pi$, $\pi$ and $1.5\pi$.
Nonetheless, our proposal is still valid for identifying the controllable MCMs because
we only need the distinct periodicity in the zero bias conductance, and we do not need any detailed and quantitative informations of the conductance spectrum.

\section{Conclusion}
\label{sec:5}
We have demonstrated that a two-dimensional noncentrosymmetric ($s+p$)-wave superconductor in the presence of an in-plane magnetic field
can host tunable Majorana corner modes (MCMs).
We also show that the number of MCMs depends sensitively on the edge shape of the noncentrosymmetric superconductor. 
We find that with irregular edge shapes there are no stable pairs of MCMs that can fully circle around the edges of the superconductor, a property that is important for braiding. 
We have also studied the effects of edge roughness on the MCMs and found that the MCMs are in general robust to edge roughness.
However, edge roughness suppresses the edge gap and may cause undesirable decoherence effects during the braiding process.
As a result, although the MCMs can be found insensitive to the edge configurations, the edge imperfections may disturb the realization of topological computations.
Therefore, developing practical measures to stabilize the braiding process of tunable MCMs, as well as fabricating clean devices,
would be the important future tasks for realizing topological quantum computations.
In addition, we discuss the tunneling spectroscopy in the presence of the MCMs, where a normal lead wire is attached to a corner of the noncentrosymmetric superconductor.
The periodicity in the zero-bias conductance with respect to the magnetic field direction is a smoking-gun signature for of the existence of controllable MCMs.
The proposed experiment is promising for observing the tunable MCMs,
because we can expect the periodicity in the zero-bias conductance to be insensitive to the details of the model.
In conclusion, we have provided groundwork knowledge for designing specific experiments for detecting tunable MCMs,
which is an essential step for realizing future topological quantum computations using higher-order topological superconductors.

\begin{acknowledgments}
This work was supported by the JSPS Core-to-Core program ``Oxide Superspin" international network.
This research was supported in part by the National Science Foundation under Grant No. NSF PHY-1748958.
\end{acknowledgments}

\appendix
\section{Bogoliubov-de Genne Hamiltonian on a tight-binding model}
\label{sec:appA}
In this section, we show the explicit form of the BdG Hamiltonian on the tight-binding model.
A lattice site is indicated by $\boldsymbol{r}=j \boldsymbol{x}+ m \boldsymbol{y}$
with $\boldsymbol{x}$($\boldsymbol{y}$) representing the unit vector in the $x$ ($y$) direction.
We assume that the superconductor is occupies for $ 1 \leq j \leq L_S$ and $1 \leq m \leq W_S$, and the hard-wall boundary condition is applied in both $x$ and $y$ directions.
The BdG Hamiltonian reads $ H = H_N + H_{\Delta}$ with
\begin{widetext}
\begin{align}
H_N = & -t \sum_{\sigma = \uparrow, \downarrow} \left[
\sum_{j=1}^{L_S-1} \sum_{m=1}^{W_S} 
\left\{c^{\dagger}_{\boldsymbol{r}+\boldsymbol{x},\sigma}c_{\boldsymbol{r},\sigma}
+c^{\dagger}_{\boldsymbol{r},\sigma} c_{\boldsymbol{r}+\boldsymbol{x},\sigma} \right\}
+ \sum_{j=+1}^{L_S} \sum_{m=1}^{W_S-1} 
\left\{c^{\dagger}_{\boldsymbol{r}+\boldsymbol{y},\sigma}c_{\boldsymbol{r},\sigma}
+c^{\dagger}_{\boldsymbol{r},\sigma} c_{\boldsymbol{r}+\boldsymbol{y},\sigma} \right\} \right] \nonumber\\
&+\sum_{j=1}^{L_S} \sum_{m=1}^{W_S}\sum_{\sigma} (4t-\mu) c^{\dagger}_{\boldsymbol{r},\sigma}c_{\boldsymbol{r},\sigma} \nonumber\\
&+\frac{i \lambda}{2} \sum_{\sigma,\sigma^{\prime}} \left[
 \sum_{j=+1}^{L_S-1} \sum_{m=1}^{W_S}
\left( \hat{\sigma} \right)_{\sigma,\sigma^{\prime}}
\left\{c^{\dagger}_{\boldsymbol{r}+\boldsymbol{x},\sigma}c_{\boldsymbol{r},\sigma^{\prime}}
-c^{\dagger}_{\boldsymbol{r},\sigma} c_{\boldsymbol{r}+\boldsymbol{x},\sigma^{\prime}} \right\}
-\sum_{j=+1}^{L_S} \sum_{m=1}^{W_S-1}\left( \hat{\sigma}_x \right)_{\sigma,\sigma^{\prime}}
\left\{c^{\dagger}_{\boldsymbol{r}+\boldsymbol{y},\sigma}c_{\boldsymbol{r},\sigma^{\prime}}
-c^{\dagger}_{\boldsymbol{r},\sigma} c_{\boldsymbol{r}+\boldsymbol{y},\sigma^{\prime}} \right\} \right] \nonumber\\
&+\sum_{j=1}^{L_S} \sum_{m=1}^{W_S} \sum_{\sigma, \sigma^{\prime}}
 \left( \boldsymbol{V}_Z \cdot \hat{\boldsymbol{\sigma}}\right)_{\sigma,\sigma^{\prime}}
c^{\dagger}_{\boldsymbol{r},\sigma}c_{\boldsymbol{r},\sigma^{\prime}},
\end{align}
and
\begin{align}
H_{\Delta} = 
&\frac{i \Delta_t}{4} \sum_{\sigma,\sigma^{\prime}} \left[
-i \sum_{j=+1}^{L_S-1} \sum_{m=1}^{W_S}
\delta_{\sigma,\sigma^{\prime}} \left\{c^{\dagger}_{\boldsymbol{r}+\boldsymbol{x},\sigma}c^{\dagger}_{\boldsymbol{r},\sigma^{\prime}}
-c^{\dagger}_{\boldsymbol{r},\sigma} c^{\dagger}_{\boldsymbol{r}+\boldsymbol{x},\sigma^{\prime}} \right\}
-\sum_{j=+1}^{L_S} \sum_{m=1}^{W_S-1}\left( \hat{\sigma}_z \right)_{\sigma,\sigma^{\prime}}
\left\{c^{\dagger}_{\boldsymbol{r}+\boldsymbol{y},\sigma}c^{\dagger}_{\boldsymbol{r},\sigma^{\prime}}
-c^{\dagger}_{\boldsymbol{r},\sigma} c^{\dagger}_{\boldsymbol{r}+\boldsymbol{y},\sigma^{\prime}} \right\} \right] \nonumber\\
&+\sum_{j=1}^{L_S} \sum_{m=1}^{W_S} \Delta_s c^{\dagger}_{\boldsymbol{r},\uparrow}c^{\dagger}_{\boldsymbol{r},\downarrow} + \mathrm{H.c.},
\end{align}
\end{widetext}
where $c^{\dagger}_{\boldsymbol{r},\sigma}$ ($c_{\boldsymbol{r},\sigma}$) is the creation (annihilation) operator of an electron
at $\boldsymbol{r}$ with spin $\sigma=\uparrow$, $\downarrow$,
$t$ denotes the nearest-neighbor hopping integral, $\mu$ is the chemical potential.
The strength of the Rashba spin-orbit coupling is represented by $\lambda$.
The Zeeman potential due to the externally applied in-plane magnetic field is given by $\boldsymbol{V}_Z = (V_Z \cos \theta_Z, V_Z \sin \theta_Z, 0)$.
The spin-singlet $s$-wave and spin-triplet $p$-wave pair potentials are denoted with $\Delta_s$ and $\Delta_t$, respectively.
The Pauli matrices in spin space are given by $\hat{\boldsymbol{\sigma}} = (\hat{\sigma}_x, \hat{\sigma}_y, \hat{\sigma}_z)$.
The normal lead wire in Fig.~\ref{fig:figure9} is described by $H_N$ with replacing $L_S \rightarrow -\infty$, $W_S \rightarrow W_N$ and $\lambda \rightarrow 0$.
The coupling between the lead wire and the superconductor is described by
\begin{align}
H_{\mathrm{int}} = -t_{\mathrm{int}} \sum_{m=1}^{W_N} \sum_{\sigma}
\left[c^{\dagger}_{j=1,m,\sigma}c_{0,m,\sigma} +c^{\dagger}_{0,m,\sigma} c_{1,m,\sigma} \right],
\end{align}
where $t_{\mathrm{int}}$ denotes the hopping integral at the junction interface.

\end{document}